\documentclass[pdftex,twocolumn,epjc3]{svjour3}          

\RequirePackage[T1]{fontenc}

\smartqed  

\RequirePackage{graphicx}
\RequirePackage{mathptmx}      
\RequirePackage{flushend}
\RequirePackage[numbers,sort&compress]{natbib}
\RequirePackage[colorlinks,citecolor=blue,urlcolor=blue,linkcolor=blue]{hyperref}

\newcommand{\orcid}[1]{\href{https://orcid.org/#1}{\,\includegraphics[width=8px]{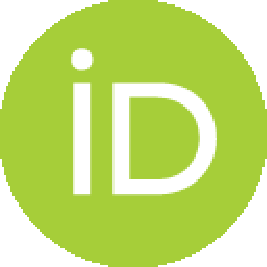}}}

\usepackage{amsmath,amssymb}

\journalname{Eur. Phys. J. C}

\begin{document}

\title{Constraints on the speed of sound in the k-essence model of dark energy}


\author{Bikash R. Dinda\orcid{0000-0001-5432-667X}\thanksref{e1,addr1}
        \and
        Narayan Banerjee \orcid{0000-0002-9799-2813}\thanksref{e2,addr1} 
}

\thankstext{e1}{e-mail: bikashd18@gmail.com}
\thankstext{e2}{e-mail: narayan@iiserkol.ac.in}

\institute{Department of Physical Sciences, Indian Institute of Science Education and Research Kolkata, Mohanpur, Nadia, West Bengal 741246, India\label{addr1}
}

\date{Received: date / Accepted: date}

\maketitle

\begin{abstract}
We consider a k-essence scalar field model for the late-time cosmic acceleration in which the sound speed, parametrized as $c_s$ is constant. We compute the relevant background and perturbation quantities corresponding to the observables like cosmic microwave background, type Ia supernova, cosmic chronometers, baryon acoustic oscillations, and the $f\sigma_8$. We put constraints on the $c_s^2$ parameter from these observations along with other parameters. We find lower values of $c_s^2$ which are close to zero are tightly constrained. Particularly, we find mean value of $\log_{\rm 10} (c_s^2)$ to be $-0.61$ and $c_s^2 \leq 10^{-3}$ is more than 3$\sigma$ away from this mean value. This means these observations favor a homogeneous dark energy component compared to the clustering one.
\end{abstract}

\section{Introduction}

The late time cosmic acceleration is confirmed by the several cosmological observations like the type Ia supernova observations \cite{SupernovaCosmologyProject:1997zqe,SupernovaSearchTeam:1998fmf,SupernovaCosmologyProject:1998vns,2011NatPh...7Q.833W,Linden2009CosmologicalPE,Camarena:2019rmj,Pan-STARRS1:2017jku,Camlibel:2020xbn}, 
cosmic microwave background (CMB) observations \cite{Planck:2013pxb,Planck:2015fie,Planck:2018vyg}, and the baryon acoustic oscillation observations \cite{BOSS:2016wmc,eBOSS:2020yzd,Hou:2020rse}. Ever since the discovery of the late time cosmic acceleration, an enormous amount of effort has been given to model this phenomenon. The two broad categories of this effort are the notion of the existence of an exotic matter called the dark energy \cite{Peebles:2002gy,Copeland:2006wr,Yoo:2012ug,Lonappan:2017lzt,Dinda:2017swh,Dinda:2018uwm} and the modification of 
gravity\cite{Clifton:2011jh,Koyama:2015vza,Tsujikawa:2010zza,Joyce:2016vqv,Dinda:2017lpz,Dinda:2018eyt,Zhang:2020qkd,Dinda:2022ixi,Bassi:2023vaq,Nojiri:2010wj,Nojiri:2017ncd,Bamba:2012cp,Lee:2022cyh}. In the first case, the dark energy is assumed to have large negative pressure which causes the late time cosmic acceleration.

There are several dark energy models in the literature \cite{Lonappan:2017lzt}. The most simple dark energy model is the $\Lambda$CDM model, in which the cosmological constant $\Lambda$ is the candidate for the dark energy 
whose energy density is considered to be a constant \cite{Carroll:2000fy}. This is the most successful model till now which can explain late-time cosmic acceleration. However, this model has some shortcomings, both from 
theoretical background like the cosmic coincidence and fine-tuning problems \cite{Zlatev:1998tr,Sahni:1999gb,Velten:2014nra,Malquarti:2003hn} and the observations point of view like corresponding to the Hubble 
tension\cite{DiValentino:2021izs,Krishnan:2021dyb,Vagnozzi:2019ezj,Dinda:2021ffa} and the $\sigma_8$ tension \cite{DiValentino:2020vvd,Abdalla:2022yfr,Douspis:2018xlj,Bhattacharyya:2018fwb}. It is thus important to study 
the cosmic acceleration with the models beyond the $\Lambda$CDM.

Different dark energy models affect cosmological evolution differently. If we assume the dark energy is homogeneous, it only affects the cosmological evolution through the background expansion and it does 
not participate in the clustering. This is the case for the $\Lambda$CDM model. However, there is no a priori reason to consider dark energy to be homogeneous. On the other hand, the inhomogeneous dark energy participates in the clustering. Hence the evolution of perturbations is different in inhomogeneous dark energy compared to the homogeneous 
one \cite{PhysRevD.81.103513,Batista:2021uhb}. Thus, it is required to check whether a dark energy is homogeneous or not.

For this purpose, we consider one popular class of dark energy models, named the k-essence \cite{Armendariz-Picon:2000ulo,Armendariz-Picon:2000nqq,Armendariz-Picon:1999hyi,Yang:2013aaa,Cardenas:2015tva,Mukherjee:2016psp,Chakraborty:2019swx,Gannouji:2020kas,Perkovic:2020eju,Huang:2021urp,Chatterjee:2022uyw,Odintsov:2020qyw,Nojiri:2019dqc}. In k-essence model, the late time acceleration is caused by a generic scalar field whose kinetic term can be both canonical and non-canonical. The canonical kinetic term corresponding to a subclass called quintessence \cite{PhysRevD.37.3406,Liddle:1998xm,Steinhardt:1999nw,Caldwell:2005tm,Scherrer:2007pu,Dinda:2016ibo}. In the quintessence model of dark energy, the speed of sound is unity. For this case, the perturbation in the scalar field is negligible. Consequently, this corresponds to the homogeneity of the dark energy. In the non-canonical k-essence scenario, the sound speed of dark energy is different from unity and it can have values lower than $1$. If the sound speed of dark energy decreases from $1$, the inhomogeneities in the dark energy may increase. For a nice review, see \cite{PhysRevD.81.103513} (also see \cite{Bamba:2011ih,Matsumoto:2014wca,Dinda:2018ojk}).

In general, any non-canonical k-essence model has an evolving speed of sound. One has to choose a model in such a way that the sound speed is always subluminal because we do not expect any information to propagate faster than the speed of light in a vacuum. Also, the sound speed has to be real. One of the easy ways to maintain such conditions is to choose a non-canonical kinetic term such that the sound speed is constant and can be parameterized. The non-canonical k-essence model with constant sound speed has been studied in the literature like in \cite{Sergijenko:2014pwa,Kunz:2015oqa,Bouhmadi-Lopez:2016cja}.

In \cite{Sergijenko:2014pwa}, authors have used Planck-2013 results on CMB anisotropy and other cosmological data to put constraints on the speed of sound and found no preferences of a particular value of it in the range between $0$ to $1$. In \cite{Kunz:2015oqa}, authors did a similar kind of analysis but with more recent data sets and found similar kinds of results. A few similar kinds of analyses have been done in the literature but with different dark energy models \cite{Hannestad:2005ak,Majerotto:2015bra,Xia:2007km}. In this study, we consider such a k-essence model in which the speed of sound is constant during the entire cosmological evolution. With this model, we study the effect of the sound speed of dark energy both in the background and the perturbation evolutions and put constraints on the sound speed of dark energy from the recent cosmological data.

Throughout our study, we consider the signature of the metric to be ($+,-,-,-$) and we quote all the expressions in the natural units. This paper is organized as follows. In Sec.~\ref{sec-Lagrangian}, we show the derivation of the form of the Lagrangian for which the speed of sound is constant over cosmic time in the k-essence scenario. In Sec.~\ref{sec-background}, we investigate the k-essence field evolution and the relevant background quantities. In Sec.~\ref{sec-perturbation}, we find the evolution of the perturbations with the full relativistic perturbation method. In Sec.~\ref{sec-autonomous}, we rewrite all the relevant background and perturbation equations in a single autonomous system of differential equations. In Sec.~\ref{sec-Newtonian}, we consider the sub-Hubble limit for the evolution equation for the matter overdensity contrast and compare the result with the full relativistic result. In Sec.~\ref{sec-data}, we briefly mention some observational data, we consider in our analysis. In Sec.~\ref{sec-result}, we discuss the results of this study. Finally, in Sec.~\ref{sec-conclusion}, we present a conclusion.

\section{K-essence Lagrangian with constant speed of sound}
\label{sec-Lagrangian}

The Lagrangian for a general k-essence scalar field, $\phi$ is given as \cite{Malquarti:2003nn,Chimento:2003zf,Pedro_Jorge_2007}

\begin{equation}
\mathcal{L}_K = \mathcal{L}_K(\phi,X),
\label{eq:general_k_essence_Lagrangian}
\end{equation}

\noindent
where $X = \frac{1}{2} (\nabla_{\mu} \phi) (\nabla^{\mu} \phi)= \frac{1}{2} (\partial_{\mu} \phi) (\partial^{\mu} \phi)$. The pressure ($P_{\phi}$), the energy density ($\rho_{\phi}$) and the sound speed ($c_s^2$) for the k-essence scalar field are given as \cite{Armendariz-Picon:1999hyi,Malquarti:2003nn,Chimento:2003zf,Pedro_Jorge_2007}

\begin{eqnarray}
P_{\phi} &=& \mathcal{L}_K,
\label{eq:P} \\
\rho_{\phi} &=& 2 X \frac{\partial \mathcal{L}_K}{\partial X}-\mathcal{L}_K,
\label{eq:rho} \\
c_s^2 &=& \frac{ \frac{\partial P_{\phi}}{\partial X} }{ \frac{\partial \rho_{\phi}}{\partial X} } = \frac{ \frac{\partial \mathcal{L}_K}{\partial X} }{ 2 X \frac{\partial^2 \mathcal{L}_K}{\partial X^2} + \frac{\partial \mathcal{L}_K}{\partial X} },
\label{eq:cs2}
\end{eqnarray}

\noindent
respectively.

We consider a k-essence model in which the speed of sound of the scalar field is constant i.e. $c_s^2=$ constant. For this case, from Eq.~\eqref{eq:cs2}, we get a differential equation for the Lagrangian given as

\begin{equation}
2 X \frac{\partial^2 \mathcal{L}_K}{\partial X^2} - \left( \frac{1-c_s^2}{c_s^2} \right) \frac{\partial \mathcal{L}_K}{\partial X} = 0,
\label{eq:L_diff_cs2_constant}
\end{equation}

\noindent
for $c_s^2 \neq 0$. The general solution for the above differential equation is given as \cite{Sergijenko:2014pwa,Kunz:2015oqa,Bouhmadi-Lopez:2016cja}

\begin{equation}
\mathcal{L}_K = U(\phi)X^n-V(\phi),
\label{eq:general_solution_L_cs2_constant}
\end{equation}

\noindent
where $U$ and $V$ are two arbitrary functions of $\phi$; and $n$ is given as

\begin{equation}
n = \frac{1+c_s^2}{2 c_s^2} = \text{constant}.
\label{eq:n_wrt_cs2}
\end{equation}

\noindent
For the simplicity of the study, we consider a special case where $U(\phi)=1$ and we denote the corresponding Lagrangian as $\mathcal{L}$ given as

\begin{equation}
\mathcal{L} = X^n-V(\phi).
\label{eq:special_L_cs2_constant}
\end{equation}

\noindent
We stick to this model throughout this study. Even though this Lagrangian is not of the standard canonical form, $V(\phi)$ can be considered as the potential for the scalar field. Eq.~\eqref{eq:n_wrt_cs2} is alternatively written as

\begin{equation}
c_s^2 = \frac{1}{2n-1} = \text{constant}.
\label{eq:cs2_wrt_n}
\end{equation}

\noindent
As we discussed in the introduction, the speed of sound should satisfy the condition $0<c_s^2\leq 1$. This corresponds to $n \geq 1$. Note that, the special case of this model, described by the lagrangian in Eq.~\eqref{eq:special_L_cs2_constant}, is the quintessence, where $n=1$ and consequently $c_s^2=1$. For other cases, $c_s^2$ decreases from the value, $1$ with increasing values of $n$.

In this model, the pressure $P_{\phi}$ of the scalar field is the same as the Lagrangian in Eq.~\eqref{eq:special_L_cs2_constant} i.e.

\begin{equation}
P_{\phi} = X^n-V(\phi).
\label{eq:P_general}
\end{equation}

\noindent
The energy density of the scalar field is given as

\begin{equation}
\rho_{\phi} = (2n-1)X^n+V(\phi).
\label{eq:rho_general}
\end{equation}

\noindent
For this model, the energy-momentum tensor, $T^{\mu}_{\nu}$ is given as

\begin{eqnarray}
T^{\mu}_{\nu} &=& \frac{\partial \mathcal{L}}{\partial (\partial _{\mu} \phi)} (\partial _{\nu} \phi) -\delta ^{\mu}_{\nu} \mathcal{L} \nonumber\\
&=& n X^{n-1} g^{\mu \sigma} (\partial_{\sigma}\phi) (\partial_{\nu}\phi) - \delta^{\mu}_{\nu} \mathcal{L},
\label{eq:Tmunu_general}
\end{eqnarray}

\noindent
where the metric of the space-time is denoted as $g_{\rm \mu \nu}$ and $\delta^{\mu}_{\nu}$ is the usual Kronecker-delta symbol.

The Euler-Lagrange equation is given as

\begin{equation}
\frac{1}{\sqrt{-g}} \partial_{\mu} \left[ \sqrt{-g} \frac{\partial \mathcal{L}}{\partial (\partial _{\mu} \phi)} \right] - \frac{\partial \mathcal{L}}{\partial \phi} = 0,
\label{eq:EL_general_pre}
\end{equation}

\noindent
where $g$ is the determinant of the given metric, $g_{\rm \mu \nu}$. In this model, the above equation consequently becomes

\begin{equation}
\frac{1}{\sqrt{-g}} \partial_{\mu} \left[ \sqrt{-g} n X^{n-1} g^{\mu \nu} (\partial_{\nu}\phi) \right] + V'(\phi) = 0.
\label{eq:EL_general}
\end{equation}

\noindent
where $V'(\phi)=\frac{d V(\phi )}{d\phi}$ and $|g|$ is the modulus of the determinant, $g$. Note that, in the above equation, we have assumed that there is no interaction between the scalar field and any other fields.

\section{Background cosmology}
\label{sec-background}

For the background cosmology, we consider the spatially flat Friedmann-Lema\^itre-Robertson-Walker (FLRW) metric given as $dS^2=dt^2-a^2(t) d\vec{r}.d\vec{r}$, where $dS$ is the line element of the space-time, $d\vec{r}$ is the comoving line element vector corresponding to the 3-dimensional Euclidean space, $t$ is the cosmic time, and $a$ is the cosmic scale factor. In this case, the expression of $X$ is given as $X=\frac{1}{2} \dot{ \bar{\phi} }^2$, where overhead dot represents the differention w.r.t $t$. Here, we denote the background scalar field as $\bar{\phi}$. Throughout this paper, a quantity with an overhead bar indicates its unperturbed (background) value. So, the background pressure ($\bar{P}_{\phi}$), energy density ($\bar{\rho}_{\phi}$), and equation of state of dark energy ($w_{\phi}$) are given as

\begin{eqnarray}
\bar{P}_{\phi} &=& \left( \frac{1}{2} \dot{\bar{\phi}}^2 \right)^n-V(\bar{\phi}), \\
\label{eq:P_special_bkg}
\bar{\rho}_{\phi} &=& (2n-1)\left( \frac{1}{2} \dot{\bar{\phi}}^2 \right)^n+V(\bar{\phi}), \\
\label{eq:rho_special_bkg}
w_{\phi} &=& \frac{\bar{P}_{\phi}}{\bar{\rho}_{\phi}} = \frac{ \left( \frac{1}{2} \dot{\bar{\phi}}^2 \right)^n-V(\bar{\phi}) }{ (2n-1)\left( \frac{1}{2} \dot{\bar{\phi}}^2 \right)^n+V(\bar{\phi}) },
\label{eq:eos_special_bkg}
\end{eqnarray}

\noindent
respectively.

The background Euler-Lagrange equation corresponding to Eq.~\eqref{eq:EL_general} is given as

\begin{equation}
n \left( \frac{1}{2} \dot{\bar{\phi}}^2 \right)^{n-1} \left[ (2 n-1) \ddot{\bar{\phi}} +3 H \dot{\bar{\phi}} \right] + V'(\bar{\phi}) = 0.
\label{eq:EL_eqn_bkg}
\end{equation}

The two Friedmann equations are given as

\begin{eqnarray}
3 M_{\rm pl}^2 H^2 &=& \bar{\rho}_{\phi} + \bar{\rho}_{m},
\label{eq:Friedmann_bkg} \\
6 M_{\rm pl}^2 (\dot{H}+H^2) &=& -(1+3 w_{\phi})\bar{\rho}_{\phi} - \bar{\rho}_{m},
\label{eq:Friedmann_bkg_2}
\end{eqnarray}

\noindent
where, $\bar{\rho}_{m}$ is the background energy density for the total matter components (including both dark matter and baryons), $H$ is the Hubble parameter, and $M_{\rm pl}^2 =\frac{1}{8 \pi G}$ with $G$ is the Newtonian gravitational constant. Here, we have neglected the radiation, because we are studying the expansion history of the Universe from the matter-dominated era to the present epoch.

\subsection{Relevant background quantities}

The energy density parameter $\Omega_{\phi}$ of the scalar field is given as

\begin{equation}
\Omega_{\phi} = \frac{\bar{\rho}_{\phi}}{3 M_{\rm pl}^2 H^2} = 1-\Omega_m,
\label{eq:bkg_OP_defn}
\end{equation}

\noindent
where $\Omega_{m}$ is the matter-energy density parameter given as

\begin{equation}
\Omega_{m} = \frac{\bar{\rho}_m}{3 M_{\rm pl}^2 H^2} = \frac{\bar{\rho}_{m0} (1+z)^3}{3 M_{\rm pl}^2 H^2} = \frac{\Omega_{\rm m0}(1+z)^3}{E^2},
\label{eq:bkg_OM}
\end{equation}

\noindent
where $\bar{\rho}_{m0}$ is the present value of the matter energy density; $\Omega_{\rm m0}$ is the present value of the matter-energy density parameter defined as $\frac{\bar{\rho}_{\rm m0}}{3 M_{\rm pl}^2 H_0^2}$ with $H_0$ being the present value of the Hubble parameter; $E$ is the normalized Hubble parameter defined as $E=\frac{H}{H_0}$. Here, we have neglected the contribution of radiation. Also, we have assumed that there is no interaction between the scalar field and the matter. The above equation can be rewritten as

\begin{equation}
E^2 = \frac{\Omega_{\rm m0}(1+z)^3}{\Omega_m}.
\label{eq:bkg_E_sqr}
\end{equation}

It is also important to calculate the cosmological distances like the luminosity distance. To do this, we define a quantity, $d_N$ given as

\begin{equation}
d_N = \int_{0}^{z} \frac{d\tilde{z}}{E(\tilde{z})},
\label{eq:defn_dN}
\end{equation}

\noindent
where $z$ (also $\tilde{z}$) is the cosmological redshift. The luminosity distance, $d_L$ and the angular diameter distance, $d_A$ is related to $d_N$ given as

\begin{eqnarray}
d_L &=& \left( \frac{1}{H_0} \right) (1+z) d_N, \\
\label{eq:dL}
d_A &=& \left( \frac{1}{H_0} \right) \frac{d_N}{1+z}.
\label{eq:dA}
\end{eqnarray}

\noindent
Note that, the above expressions are valid for the spatially flat Universe assumption which we consider throughout our analysis.

\section{Evolution of perturbations}
\label{sec-perturbation}

In this analysis, we are interested in the linear perturbations for the evolutions of scalar fluctuations only for which we can compute the evolution of the scalar fluctuations independently with the two scalar degrees of freedom. Further, we assume there is no source of anisotropic stress. So, we can compute the evolution of the scalar fluctuations with only one scalar degree of freedom. Here, we are considering the conformal Newtonian gauge for which the perturbed metric is given as \cite{Dinda:2017lpz,Dinda:2016ibo}

\begin{equation}
dS^2 = (1+2 \Phi)dt^2-a^2(1-2\Phi) d\vec{r}.d\vec{r},
\label{eq:pert_metric}
\end{equation}

\noindent
where $\Phi$ is the gravitational potential or sometimes it is called the Bardeen potential. With the above metric, the first-order Euler-Lagrange equation corresponding to Eq.~\eqref{eq:EL_general} becomes

\begin{eqnarray}
&& 2^{1-n} n \left(\dot{ \bar{\phi} }^2\right)^n \Bigg{[} (2n-1) \ddot{(\delta \phi)} + 3 H \dot{(\delta \phi)} - 2 (n+1) \dot{ \bar{\phi} } \dot{\Phi } \nonumber\\
&& - \frac{1}{a^2} \nabla^2(\delta \phi) \Bigg{]} + 2 \dot{ \bar{\phi} } V'(\bar{\phi}) \left[ (1-n)\dot{(\delta \phi)} +n \dot{ \bar{\phi} } \Phi \right] \nonumber\\
&& + \dot{ \bar{\phi} }^2 V''(\bar{\phi}) (\delta \phi) = 0,
\label{eq:first_order_EL_eqn}
\end{eqnarray}

\noindent
where we have perturbed the scalar field as $\phi = \bar{\phi}+\delta \phi$ and $k$ is the magnitude of wave vector. The above equation is obtained from the Fourier transform of the first-order perturbations. Throughout this study, we mention all the first-order perturbation equations in the Fourier space. The first-order field equations are given as

\begin{align}
\label{eq:first_order_field_eqns_1}
& \nabla^2\Phi - 3a^2H(\dot{\Phi}+H\Phi) = 4\pi Ga^2(\delta \rho_{\phi}+\bar{\rho}_{m}\delta_{m}), \\
\label{eq:first_order_field_eqns_2}
& \dot{\Phi}+H\Phi = 4\pi G \left[ a (\bar{\rho}_{\phi}+\bar{P}_{\phi})v_{\phi} + a \bar{\rho}_{m} v_m \right], \\
\label{eq:first_order_field_eqns_3}
& \ddot{\Phi}+4H\dot{\Phi}+(2\dot{H}+3H^2)\Phi = 4\pi G (\delta P_{\phi}),
\end{align}

\noindent
where $\bar{\rho}_{m}\delta_{m}$ is the perturbation in the energy density of the matter components, $\bar{\rho}_{m}$ is the background matter energy density and $\delta_m$ is the matter overdensity contrast, defined as $\delta_m = \frac{\rho_{m}-\bar{\rho}_{m}}{\bar{\rho}_{m}}$, where $\rho_{m}$ is the total matter-energy density. $v_m$ is the velocity perturbations in the matter fields. $\delta \rho_{\phi}$, $v_{\phi}$ and $\delta P_{\phi}$ are the first-order perturbations in the energy density, velocity field, and pressure respectively for the scalar field. These are given as

\begin{align}
\label{eq:first_order_rho_phi}
& \delta \rho_{\phi} = 2^{1-n} n (2 n-1) \left(\dot{ \bar{\phi} }\right)^{2n-1} \left[ \dot{(\delta \phi)}- \dot{ \bar{\phi} } \Phi \right] \nonumber\\
& + V'(\bar{\phi}) \delta \phi, \\
\label{eq:first_order_v_phi}
& a (\bar{\rho}_{\phi}+\bar{P}_{\phi})v_{\phi} = 2^{1-n} n \left(\dot{ \bar{\phi} }\right)^{2n-1} \delta \phi, \\
\label{eq:first_order_P_phi}
& \delta P_{\phi} = 2^{1-n} n \left(\dot{ \bar{\phi} }\right)^{2n-1} \left[ \dot{(\delta \phi)}- \dot{ \bar{\phi} } \Phi \right] \nonumber\\
& - V'(\bar{\phi}) \delta \phi.
\end{align}

\noindent
In general, we need to solve Eqs.~\eqref{eq:first_order_EL_eqn},~\eqref{eq:first_order_field_eqns_1},~\eqref{eq:first_order_field_eqns_2}, and~\eqref{eq:first_order_field_eqns_3} simultaneously to find the solutions 
for $\delta \phi$, $\Phi$, $\delta _m$ and $v_m$. However, the expressions for these differential equations are such that we do not need to solve all the differential equations simultaneously. Instead, we can only simultaneously 
solve Eqs.~\eqref{eq:first_order_EL_eqn} and~\eqref{eq:first_order_field_eqns_3} to find solutions for $\delta \phi$ and $\Phi$ first, because these two differential equations are in the closed form w.r.t the quantities $\delta \phi$ 
and $\Phi$ and other two quantities $\delta _m$ and $v_m$ are not explicitely present. Using the solutions of $\delta \phi$ and $\Phi$ in Eqs.~\eqref{eq:first_order_field_eqns_1} and~\eqref{eq:first_order_field_eqns_2}, $\delta _m$ and $v_m$ can be solved 
separately. Putting Eq.~\eqref{eq:first_order_P_phi} in Eq.~\eqref{eq:first_order_field_eqns_3}, we get a differential equation for $\Phi$ given as

\begin{align}
\label{eq:first_order_field_eqns_3_Again}
& \ddot{\Phi}+4H\dot{\Phi}+(2\dot{H}+3H^2)\Phi = 4\pi G \nonumber\\
& \left( 2^{1-n} n \left(\dot{ \bar{\phi} }\right)^{2n-1} \left[ \dot{(\delta \phi)}- \dot{ \bar{\phi} } \Phi \right] - V'(\bar{\phi}) \delta \phi \right).
\end{align}

\noindent
As mentioned before, since, Eqs.~\eqref{eq:first_order_EL_eqn} and~\eqref{eq:first_order_field_eqns_3_Again} are in closed form, we numerically solve these two equations simultaneously to find solutions of $\delta \phi$ and $\Phi$.

\subsection{Relevant perturbation quantities}

We put Eq.~\eqref{eq:first_order_rho_phi} in Eq.~\eqref{eq:first_order_field_eqns_1} and algebraically solve it to find expression for $\delta_m$ given as

\begin{align}
\label{eq:delm_pre}
& \delta_m = -\frac{1}{\bar{\rho}_m} \Bigg{(} \frac{- \nabla^2\Phi + 3a^2H(\dot{\Phi}+H\Phi)}{4\pi G a^2} \nonumber\\
& + 2^{1-n} n (2 n-1) \left(\dot{ \bar{\phi} }\right)^{2n-1} \left[ \dot{(\delta \phi)}- \dot{ \bar{\phi} } \Phi \right] + V'(\bar{\phi}) \delta \phi \Bigg{)}.
\end{align}

\noindent
Similarly, we put Eq.~\eqref{eq:first_order_v_phi} in Eq.~\eqref{eq:first_order_field_eqns_2} and algebraically solve it to find expression for $v_m$ given as

\begin{equation}
v_m = \frac{1}{a \bar{\rho}_m} \left[ \frac{\dot{\Phi}+H\Phi}{4\pi G} - 2^{1-n} n \left(\dot{ \bar{\phi} }\right)^{2n-1} \delta \phi \right] .
\label{eq:vm_pre}
\end{equation}

\section{Autonomous system of differential equations and initial conditions: background and perturbation togther}
\label{sec-autonomous}

From now onward we mention all the equation in the Fourier space. The equations can just simply be rewritten by replacing $\nabla^2 f$ with $-k^2 f$ for any perturbed quantity $f$ which is the function of the spatial coordinates, where $k$ is the wavenumber. And for simplicity i.e. to avoid any complicated notations, we use same notations for perturbed quantities in the Fourier space too.

\subsection{Autonomous system}

We define some dimensionless variables for the background quantities given as

\begin{eqnarray}
x &=& \frac{2^{-\frac{n}{2}} \sqrt{2 n-1} (\dot{\bar{\phi}})^n}{\sqrt{3} H M_{\rm pl}}, \nonumber\\
A &=& \frac{\sqrt{V(\bar{\phi})}}{ 2^{-\frac{n}{2}} \sqrt{2 n-1} (\dot{\bar{\phi}})^n }, \nonumber\\
\lambda &=& -\frac{2^{\frac{n-1}{2}} M_{\text{pl}} (\dot{\bar{\phi}})^{1-n} V'(\bar{\phi})}{\sqrt{2 n-1} V(\bar{\phi})}, \nonumber\\
B &=& \ln \left[ \frac{(1+z)^{-\frac{3}{2}}E}{(1+z_i)^{-\frac{3}{2}}E_i} \right], \nonumber\\
F &=& (1+z_i)^{-\frac{3}{2}}E_i(d_N-d_N^i),
\label{eq:bkg_variables_1}
\end{eqnarray}

\noindent
where, $E_i$ and $d_N^i$ are the initial values of the quantities $E$ and $d_N$ respectively, at an initial redshift, $z_i$. Similarly, we define dimensionless variables for the first-order perturbation quantities given as

\begin{eqnarray}
Q &=& \left( \frac{d \bar{\phi} }{dN} \right)^{-1} \delta \phi, \nonumber\\
R &=& \frac{dQ}{dN}, \nonumber\\
S &=& \frac{d\Phi}{dN}.
\label{eq:variables_perturbations}
\end{eqnarray}

\noindent
With these variables, the background and perturbation equations all together are written in an autonomous system of differential equations given as

\begin{align}
\label{eq:dyn_sys_full}
\frac{dx}{dN} =& \frac{x}{2} \left[ A^2 x \left(\sqrt{6} \lambda -3 x\right)+\frac{3 \left(x^2-1\right)}{2 n-1} \right], \nonumber\\
\frac{dA}{dN} =& \frac{3 A n}{2 n-1}-\sqrt{\frac{3}{2}} A \left(A^2+1\right) \lambda  x, \nonumber\\
\frac{d\lambda}{dN} =& \frac{\sqrt{\frac{3}{2}} \lambda ^2 x \left[ A^2-n \left(A^2+2 \Gamma -2\right) \right] }{n}+\frac{3 \lambda  (n-1)}{2 n-1}, \nonumber\\
\frac{dB}{dN} = & \frac{3 x^2 \left[ A^2 (2 n-1)-1 \right] }{2(2n-1)}, \nonumber\\
\frac{dF}{dN} = & -e^{\frac{N}{2}-B}, \nonumber\\
\frac{d\Phi}{dN} = & S, \nonumber\\
\frac{dQ}{dN} = & R, \nonumber\\
\frac{dR}{dN} = & f_1, \nonumber\\
\frac{dS}{dN} = & f_2,
\end{align}
\noindent
with

\begin{align}
f_1 &= \frac{1}{2(2n-1)^2} \Bigg{[} x \Bigg{(} 3 Q x \Bigg{[} \nonumber\\
& \frac{A (1-2 n)^2 \left(\lambda ^2 \left(A^2 (n-1)-2 A n+A-n\right)+3 A n^2\right)}{n^2} \nonumber\\
& +3 \Bigg{]} -9 Q x^3 \left(A^2 (1-2 n)+1\right)^2 \nonumber\\
& +3 \sqrt{6} A \lambda  (2 n-1) Q x^2 \left(-A^2+2 (A-1) A n-1\right) \nonumber\\
& -2 \sqrt{6} A \lambda  (2 n-1) (3 (n-1) Q-2 n \Phi +\Phi ) \Bigg{)} \nonumber\\
& +(2 n-1) R \Big{[} 3 x^2 \left(A^2 (2 n-1)-1\right) \nonumber\\
& +2 \sqrt{6} A \lambda  (1-2 n) x-6 n+9 \Big{]} \nonumber\\
& +2 (1-2 n) Q k_n^2+4 \left(2 n^2+n-1\right) S \Bigg{]},
\end{align}

\noindent
and

\begin{align}
f_2 &= \frac{1}{2(2n-1)^2} \Bigg{[} -3 x^2 (n^2 \Big{(} 4 \Big{(} A^2 (S+2 \Phi ) \nonumber\\
& -R+\Phi \Big{)} -6 Q \Big{)} \nonumber\\
& +n (9 Q-2 \left(2 A^2 S+4 A^2 \Phi -R+S+3 \Phi \right)) \nonumber\\
& +\left(A^2+1\right) (S+2 \Phi )) +9 n Q x^4 \left(A^2 (1-2 n)+1\right) \nonumber\\
& +6 \sqrt{6} A \lambda  n (2 n-1) Q x^3-5 (1-2 n)^2 S \Bigg{]},
\end{align}

\noindent
where $N=\ln{a}$ and $\Gamma$ is defined as

\begin{equation}
\Gamma = \frac{ V(\bar{\phi}) V''(\bar{\phi}) }{\left[ V'(\bar{\phi}) \right]^2},
\label{eq:Gamma}
\end{equation}

\noindent
where $V''(\bar{\phi})=\frac{d^2V(\bar{\phi})}{d\bar{\phi}^2}$. In the above equation, $\Gamma$ is defined in a way such that for the polynomial and the exponential potentials, $\Gamma$ becomes constant. We restrict our study to these kinds of potentials only and these would be enough to convey the results. Here, $k_n$ is defined as

\begin{eqnarray}
&& k_n = \frac{k}{a H} = \tilde{k} e^{\frac{N}{2}-B},
\label{eq:k_norm} \\
&& \text{with} \hspace{0.5 cm} \tilde{k} = \frac{k e^{B_0}}{H_0}.
\label{eq:k_tilde}
\end{eqnarray}

\noindent
To solve the system of differential equations in Eq.~\eqref{eq:dyn_sys_full}, we keep $\tilde{k}$ as a free parameter, and after obtaining the solutions we convert it to get the usual magnitude, $k$ of the wavevector using Eq.~\eqref{eq:k_tilde}. Note that, in Eq.~\eqref{eq:dyn_sys_full}, there is no variable involved in the denominators except constant factors $n$ and $2n-1$. Both these constant factors are not zero for $n \geq 1$. So, there are no singularity issues in the above system of differential equations.

\subsection{Initial conditions}

To solve the set of differential equations in Eq.~\eqref{eq:dyn_sys_full}, we need to fix the initial conditions. We denote initial values by subscript 'i' or in some cases by superscript 'i'. We fix the initial conditions at a redshift,

\begin{equation}
z_i=1100.
\end{equation}

The quantities $B$ and $F$ are defined in such a way that their initial values are

\begin{eqnarray}
B_i &=& 0, \\
F_i &=& 0,
\end{eqnarray}

\noindent
respectively. These initial values are consistent with the fact that the normalized Hubble parameter is unity and the distances are zero at the present epoch i.e. $z=0$.

The initial values, $x_i$ and $A_i$ are related to the initial values, $\gamma_{\phi}^i$ and $\Omega _{\phi }^i$ given as

\begin{eqnarray}
x_i &=& \sqrt{\gamma _{\phi }^i \left(\Omega _{\phi }^i-\frac{\Omega _{\phi }^i}{2 n}\right)}, \nonumber\\
A_i &=& \frac{\sqrt{\gamma _{\phi }^i-2 n \gamma _{\phi }^i+2 n}}{\sqrt{(2 n-1) \gamma _{\phi }^i}},
\label{eq:bkg_initial_conditions_1}
\end{eqnarray}

\noindent
where $\gamma_{\phi}^i =1+w_{\phi}^i$. These two parameters can be related to $\Omega_{\rm m0}$ and $w_0$ (equation of state of the scalar field at present). These relations are not analytic but can be computed numerically.

We keep $\lambda_i$ as a free parameter which corresponds to the initial slope of the given potential of the scalar field.

We also keep $\Gamma$ as a free parameter. That means we do not choose any specific potential. The only assumption here is that we stick to such potentials for which $\Gamma$ is a constant. This is the case for the polynomial and exponential potentials, as mentioned previously.

The initial conditions corresponding to the perturbation equations are given as \cite{Dinda:2017lpz,Dinda:2016ibo}

\begin{eqnarray}
\Phi_i &=& -\frac{3}{2} \frac{a_i^3 H_i^2}{k^2} = -\frac{3}{2 \tilde{k}^2}, \nonumber\\
Q_i &=& 0, \nonumber\\
R_i &=& 0, \nonumber\\
S_i &=& 0.
\label{eq:pert_relativistic_initials}
\end{eqnarray}

\noindent
$Q_i$ and $R_i$ are taken to be zero because in a matter-dominated era, at $z_i=1100$, there are hardly any dark energy contributions both in the background and the perturbation. In the early matter-dominated era, $\Phi$ is approximately constant. So, we choose $S_i=0$. The initial value, $\Phi_i$ is computed from the assumption that at matter dominated era, $\delta_m \propto a$ at sub-Hubble scale. So, all the parameters related to the initial conditions for the perturbed quantities are fixed.

So, in this analysis, the model parameters are $\gamma _{\phi }^i$, $\Omega _{\phi }^i$, $\lambda _i$, and $\Gamma$.

\subsection{Background quantities w.r.t dimensionless variables}

With the dimensionless variables, $x$, $A$, and $\lambda$, defined in Eq.~\eqref{eq:bkg_variables_1}, the equation of state and the energy density parameter ($\Omega_{\phi}$) of the scalar field are expressed as

\begin{eqnarray}
w_{\phi} &=& \frac{2 n}{\left(A^2+1\right) (2 n-1)}-1, \\
\label{eq:bkg_w}
\Omega_{\phi} &=& \left(1+A^2\right) x^2,
\label{eq:bkg_OP}
\end{eqnarray}

\noindent
respectively. The normalized Hubble parameter $E$ is computed from the quantity $B$ given as

\begin{equation}
E = (1+z)^{\frac{3}{2}} e^{B-B_0},
\label{eq:bkg_E_dffrnt}
\end{equation}

\noindent
where $B_0=B(z=0)$. Similarly, $d_N$ is computed from the quantity $F$ as

\begin{equation}
d_N = (F-F_0) e^{B_0},
\label{eq:bkg_dN_dffrnt}
\end{equation}

\noindent
where $F_0=F(z=0)$.

\subsection{Perturbation quantities w.r.t dimensionless variables}

We use Eq.~\eqref{eq:delm_pre} to compute the perturbation in the matter-energy density given as

\begin{align}
\label{eq:delm}
\delta_m &= \left[ \left(A^2+1\right) x^2-1 \right]^{-1} \Bigg{(} n x^2 \Bigg{[} \nonumber\\
& \frac{3 Q \left(n \left(2-2 A^2 x^2\right)+\left(A^2+1\right) x^2-3\right)}{2 n-1} \nonumber\\
& +2 (R-\Phi ) \Bigg{]} +\frac{2}{3} \Phi  \left(k_n^2+3\right)+2 S \Bigg{)}.
\end{align}

Similarly, we use Eq.~\eqref{eq:vm_pre} to compute the perturbation in the velocity field of matter as

\begin{equation}
3 a H v_m = \frac{2 \left(3 n Q x^2-2 n S-2 n \Phi +S+\Phi \right)}{(2 n-1) \left[ \left(A^2+1\right) x^2-1 \right]}.
\label{eq:ym}
\end{equation}

\noindent
Using Eqs.~\eqref{eq:delm} and~\eqref{eq:ym}, we get the gauge invariant matter energy density contrast, $\Delta_m$ given as

\begin{equation}
\Delta_m = \delta_m+3 a H v_m.
\label{eq:Deltam}
\end{equation}

\section{Sub-Hubble limit of perturbations, logarithmic growth factor, and $\sigma_8$}
\label{sec-Newtonian}

Using the non-relativistic approximations, such as $k \gg aH$ (sub-Hubble) and spatial variations like $\partial _i \Phi$ and $\nabla ^2 \Phi$ are much greater than the temporal variations $\dot{\Phi}$ or $\ddot{\Phi}$ (quasistatic) in Eqs.~\eqref{eq:first_order_EL_eqn},~\eqref{eq:first_order_field_eqns_1}, and~\eqref{eq:first_order_field_eqns_2}, one can arrive at the corresponding equation for the Newtonian perturbation theory. For a detailed discussion, see \cite{Bamba:2011ih}. The relevant equation for $\delta _m$ looks like,

\begin{equation}
\ddot{\delta}_m^N+2H\dot{\delta}_m^N-4\pi G\bar{\rho}_m\delta_m^N = 0.
\label{eq:pert_Newtonian}
\end{equation}

\noindent
So, in the sub-Hubble scale, we can solve this simple differential equation instead of solving the complicated differential equations in the relativistic perturbations. The superscript 'N' corresponds to the case of Newtonian perturbation theory. Note that, even though sound speed is not unity in this model, the nature of the scalar field is such that the parameter $c_s$ does not appear explicitly in the above differential equation. This equation is also written in a system of differential equations given as

\begin{align}
\label{eq:pert_dyn_sys_2_Newtonian}
\frac{d\delta_m^N}{dN} &= T, \nonumber\\
\frac{dT}{dN} &= \frac{1}{2} \Bigg{[} -3 \left(A^2+1\right) x^2 \delta _m^N \nonumber\\
& +3 T x^2 \left(\frac{1}{2 n-1}-A^2\right) +3 \delta _m^N-T \Bigg{]}.
\end{align}

\noindent
To solve system of differential equations in Eq.~\eqref{eq:pert_dyn_sys_2_Newtonian}, we use the usual initial conditions given as $\delta_m^N(z=z_i)=a_i$ and $T(z=z_i)=a_i$. This comes from the fact that in the early matter-dominated era, $\delta _m \propto a$ \cite{Dinda:2017swh}.

\begin{figure}
\centering
\includegraphics[width=0.45\textwidth]{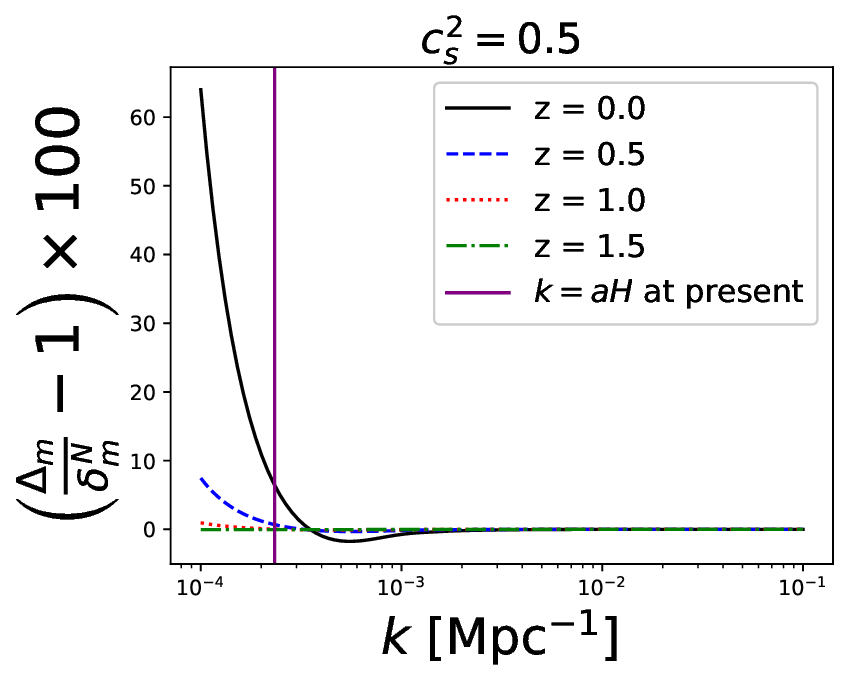}
\includegraphics[width=0.45\textwidth]{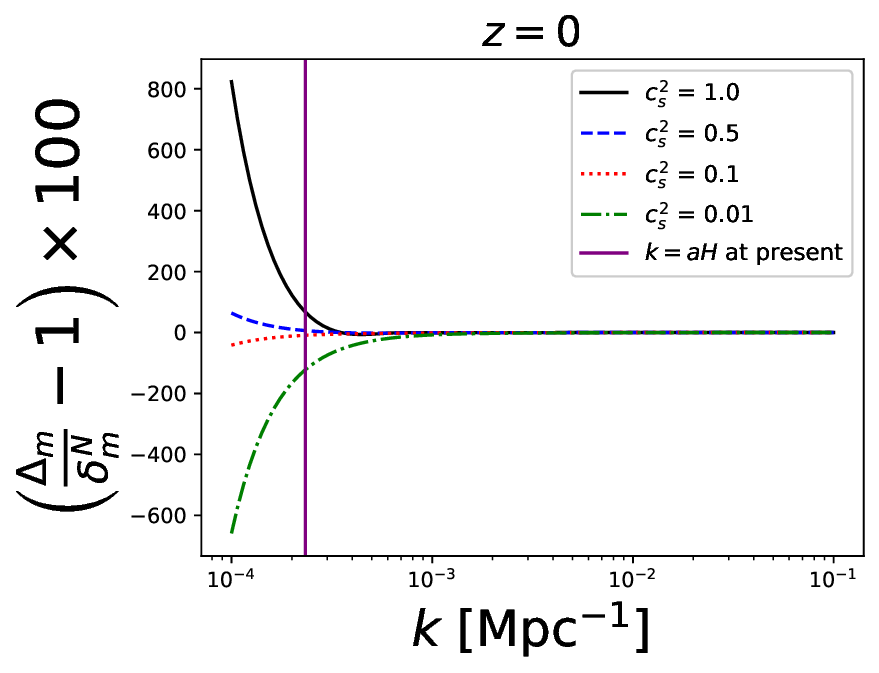}
\caption{
\label{fig:cmp_N_R}
Percentage deviation of the relativistic perturbation results for $\Delta_m$ compared to the Newtonian perturbation result for $\delta_m^N$.
}
\end{figure}

In Figure~\ref{fig:cmp_N_R}, we have compared the results from relativistic and Newtonian perturbation theory for the matter-energy overdensity contrast. The vertical purple line is for the horizon scale corresponding to  $k_H(z=0)=a(z=0)H(z=0) \approx 0.00023$ Mpc$^{-1}$ for $h=0.7$, where $h$ is related to $H_0$ given as

\begin{equation}
H_0 = 100 \hspace{0.2 cm} h \hspace{0.2 cm} \text{km} \hspace{0.1 cm} \text{s}^{-1} \hspace{0.1 cm} \text{Mpc}^{-1}.
\label{eq:defn_h}
\end{equation}

\noindent
We find that for almost all the cases the relativistic perturbation results match well within 10$\%$ with the Newtonian perturbation results for the sub-Hubble scales. We are interested in the sub-Hubble scales, so from now on we shall use the Newtonian perturbation results. The growth factor corresponding to the matter inhomogeneities is given as \cite{Huterer:2013xky}

\begin{equation}
f = \frac{d \ln D_{+} }{d \ln a} = \frac{T}{\delta_m^N},
\label{eq:growth_f}
\end{equation}

\noindent
where $D_{+}$ is the growing mode solution of $\delta_m^N$. In the second equality in the above equation and from now onwards we denote the growing mode solution with the same notation $\delta_m^N$.

In the Newtonian perturbation theory, the normalization factor of the matter power spectrum, $\sigma_8$ is independent of the scale $k$ and it is written as \cite{Pierpaoli:2000ip}

\begin{equation}
\sigma_8 = \sigma_8^0 \frac{D_{+}(z)}{D_{+}(z=0)} = \sigma_8^0 \frac{\delta_m^N(z)}{\delta_m^N(z=0)},
\label{eq:growth_f}
\end{equation}

\noindent
where $\sigma_8^0$ is the present value of $\sigma_8$.

\section{Observational data}
\label{sec-data}

We consider Planck 2018 results of the cosmic microwave background (CMB) observations for the 'TT,TE,EE +lowl +lowE +lensing' with the base flat $\Lambda$CDM model, where 'T' stands for temperature in the CMB map and 'E' stands for E-mode of the CMB polarisation map \cite{Planck:2018vyg}. For this purpose, in this analysis, we use the CMB distance prior data corresponding to these observations \cite{Zhai:2018vmm,Chen:2018dbv}. For the CMB distance prior, we use the corresponding constraints on the CMB shift parameter, acoustic length scale, and the present value of baryon energy density parameter ($\Omega_{\rm b0}$) according to the \cite{Chen:2018dbv}. We denote this observation as 'CMB' throughout this analysis.

We consider Pantheon compilation of the type Ia supernova observations which possesses apparent peak absolute magnitudes of the standard candles at different redshift values \cite{Pan-STARRS1:2017jku}. This apparent magnitude depends on the value of the luminosity of a source at a particular redshift and the nuisance parameter $M_B$. $M_B$ is the peak absolute magnitude of a type Ia supernova. We constrain $M_B$ alongside the model parameters. We denote this observation as 'SN'.

We consider the cosmic chronometer data for the Hubble parameter at different redshift values \cite{Pinho:2018unz,Jimenez:2001gg}. In these observations, the Hubble parameter is determined by the relative galaxy ages. For the Hubble parameter data, we closely follow \cite{Jimenez:2001gg}. We denote this observation as 'CC'.

We consider baryon acoustic oscillations (BAO) data which are related to the cosmological distances like the angular diameter distance. The BAO observations possess data both in the line of sight direction and transverse direction \cite{eBOSS:2020yzd}. The line of sight data is related to the Hubble parameter and the transverse data is related to the angular diameter distance \cite{BOSS:2016wmc,eBOSS:2020yzd,Hou:2020rse}. For the BAO data, we follow \cite{eBOSS:2020yzd}. However, we exclude the measurement of eBOSS (the extended baryon oscillation spectroscopic survey) emission-line galaxies (ELGs) data from the list in \cite{eBOSS:2020yzd} because this data (at redshift, $z=0.8$) have an asymmetric standard deviation in the statistical measurement. Note that BAO observation is dependent on the parameter, $r_d$, the distance to the baryon drag epoch. This parameter is closely related to the parameter, $\Omega_{\rm b0}$. So, in our analysis, we constrain this parameter as a nuisance parameter like in the case of CMB data. We denote the BAO observations as 'BAO'.

We also consider the $f\sigma_8$ data in our analysis. This data constrains the model parameters both through background and perturbation evolutions. We consider 63 $f\sigma_8$ data at different redshift ranging from $z=10^{-3}$ to $z=2$. For these data, we follow \cite{Kazantzidis:2018rnb}. We denote these observations as '$f\sigma_8$'. With all these data, we constrain the model parameters alongside the cosmological nuisance parameters.

\begin{figure*}
\centering
\includegraphics[width=1.0\textwidth]{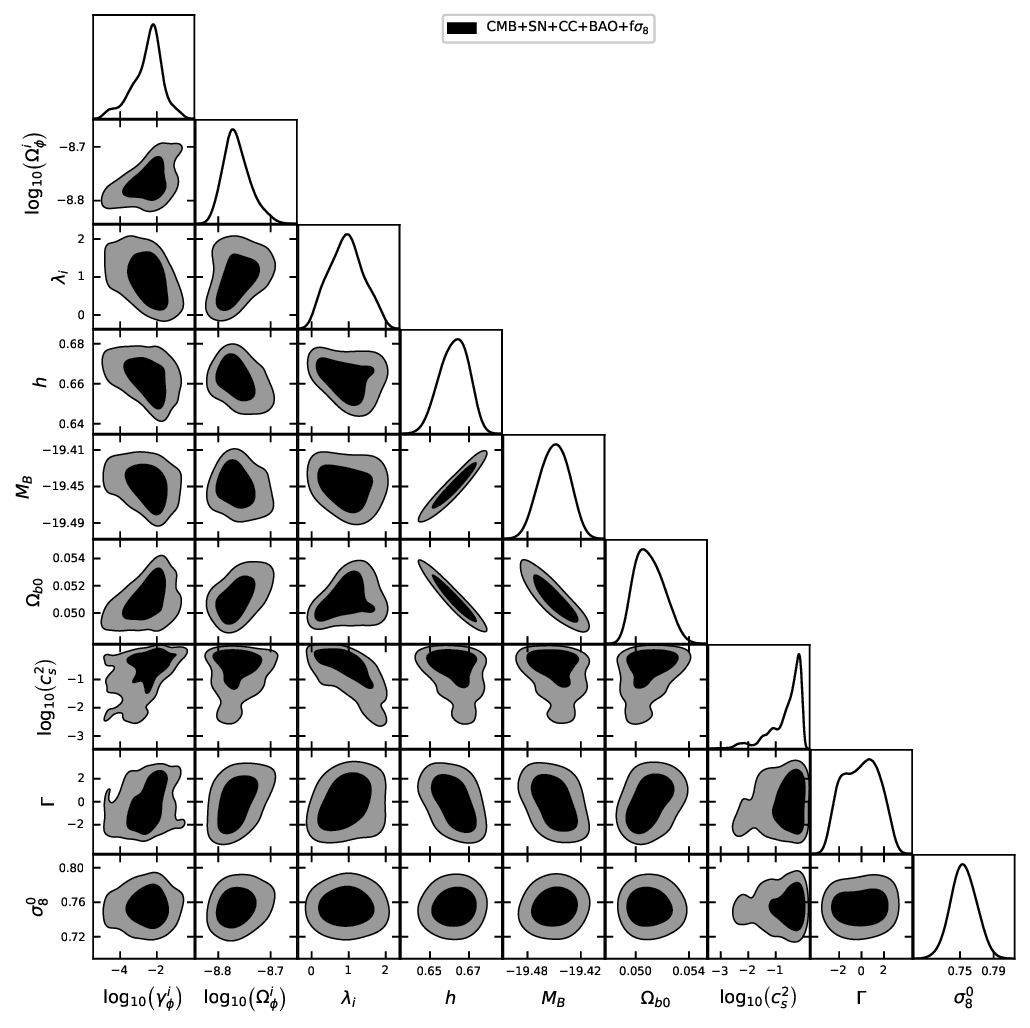}
\caption{
\label{fig:bounds_main}
Constraints on all the model parameters obtained from the CMB+SN+CC+BAO+$f\sigma_8$ combinations of data. The inner and outer regions correspond to 1$\sigma$ and 2$\sigma$ bounds respectively.
}
\end{figure*}

\begin{figure}
\centering
\includegraphics[width=0.45\textwidth]{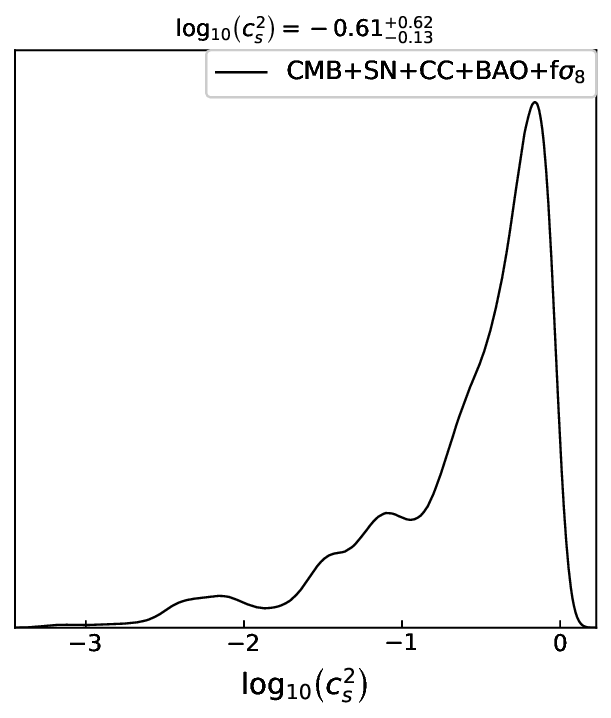}
\caption{
\label{fig:prob_log10_cs2}
Marginalized probability of $\log_{10}c_s^2$ obtained from CMB+SN+CC+BAO+$f\sigma_8$ combinations of data.
}
\end{figure}

\section{Results}
\label{sec-result}

\begin{table}
\begin{center}
\begin{tabular}{ |c|c|  }
\hline
Parameters & 1$\sigma$ bounds \\
\hline
$\log_{\rm 10}(\gamma_{\phi}^i)$ & $-2.52^{+0.88}_{-0.63}$ \\
\hline
$\log_{\rm 10}(\Omega_{\phi}^i)$ & $-8.764^{+0.020}_{-0.029}$ \\
\hline
$\lambda_i$ & $0.94^{+0.45}_{-0.53}$ \\
\hline
$h$ & $0.6624^{+0.0081}_{-0.0069}$ \\
\hline
$M_B$ & $-19.450\pm0.016$ \\
\hline
$\Omega_{\rm b0}$ & $0.05102^{+0.00095}_{-0.0013}$ \\
\hline
$\log_{\rm 10}(c_s^2)$ & $-0.61^{+0.62}_{-0.13}$ \\
\hline
$\Gamma$ & $-0.1\pm1.6$ \\
\hline
$\sigma_8^0$ & $0.754\pm0.015$ \\
\hline
\end{tabular}
\end{center}
\caption{
1$\sigma$ bounds on all the model parameters obtained from CMB+SN+CC+BAO+$f\sigma_8$ combinations of data.
}
\label{table:main}
\end{table}

In Figure~\ref{fig:bounds_main}, we have shown constraints on all the parameters obtained from the combined CMB+SN+CC+BAO+$f\sigma_8$ data. The inner-darker-black and outer-lighter-black contours correspond to the 1$\sigma$ and 2$\sigma$ contour ellipses respectively. The 1$\sigma$ values of parameters are mentioned in Table~\ref{table:main}.

As we can see, from the combinations of all the data, mentioned earlier, the higher values of $c_s^2$ (close to $1$) are not tightly constrained. But, interestingly, constraints on the lower values of $c_s^2$ (close to $0$) are tighter. This can also be seen from Figure~\ref{fig:prob_log10_cs2}, where we have shown the marginalized probability of $\log_{10}c_s^2$. This means the homogeneous dark energy is more favorable than the clustering dark energy from the recent observational data, we have considered. This analysis shows tighter constraints on $c_s^2$ (on the lower side i.e. close to $0$) compared to the results obtained in the earlier studies like in \cite{Sergijenko:2014pwa,Kunz:2015oqa,Bouhmadi-Lopez:2016cja,Hannestad:2005ak,Majerotto:2015bra,Xia:2007km}. This is our main highlighted result. However, from the constraints on the different model parameters, we see some other interesting results, mentioned below.

From the constraints on $\gamma_{\phi}^i$, we see that its mean value is of the order of $10^-2$ which corresponds to the fact that the equation of the state parameter of the dark energy is very close to $-1$ at the initial time. This means the initial condition for the scalar field evolution favors the thawing behavior for a larger set of forms of potential including polynomials and the exponential. Even the negative values of the powers in the polynomial potentials also favor the thawing behaviors which can be seen from the constraints on the $\Gamma$ parameters, in which we see the large range of the parameter space is allowed for $\Gamma$ including $0$ (corresponding to the exponential potential) and positive values (corresponding to the negative power of the polynomial potentials). Note that, these results are only for the polynomial and exponential potentials not for any arbitrary general form of potential.

The results for the constraints on the $H_0$ parameter (through the parameter, $h$) are similar to the ones, we expect from the CMB, CC, and BAO observations. Since $M_B$ is degenerate to $H_0$, constraints on $M_B$ are also consistent. The constraints on the $\Omega_{\rm b0}$ are also consistent, as we expect from the CMB and the BAO observations. Similar is the case for the constraints on the $\sigma_8^0$ parameter.

We should note that two related studies explore similar models, incorporating constraints from similar cosmological observations. In \cite{Sergijenko:2014pwa,Kunz:2015oqa}, authors considered similar k-essence models with different potentials for the scalar field. The main improvement in the present work is that we have considered an updated dataset from Planck 2018 \cite{Planck:2018vyg} compared to that used in the previous investigations \cite{Sergijenko:2014pwa,Kunz:2015oqa}, along with updated BAO \cite{eBOSS:2020yzd} and CC \cite{Jimenez:2001gg} dataset as well. This leads to a slightly tighter constraint on the results. As we are working at the values of $c_s^2$ ranging from unity to very close to zero, the use of a logarithmic scale ($\log_{\rm 10}\left(c_s^2\right)$) leads to a better distinguishability close to $c_s^2=0$ in the present work. A similar ploy was used in \cite{Kunz:2015oqa} as well, but not in \cite{Sergijenko:2014pwa}.

\section{Conclusion}
\label{sec-conclusion}

We consider a k-essence model of dark energy in which the sound speed of dark energy is constant. We write down the corresponding Lagrangian for this kind of model. With this Lagrangian, we calculate the Euler-Lagrange equation and the field equations in general. We then set up a dynamical system of differential equations for the background evolutions with the help of dimensionless variables. After numerically solving this autonomous system, we compute the relevant background quantities like the Hubble parameter, the equation of state parameter of the dark energy, and the energy density parameter of the k-essence scalar field.

We also compute the first-order linear perturbations to compute the relevant perturbation quantities like the growth factor and the $\sigma_8$. We consider both the relativistic and the Newtonian perturbations and compare them. We find the results match excellent within the sub-Hubble limit. For the evolution of the perturbations, we use dimensionless variables to get an autonomous system of differential equations. We combine this autonomous system with the one for background evolution and make a completely autonomous system of differential equations. From this, we compute all the relevant quantities.

Next, we do the parameter estimation to put constraints on the model parameters as well as on the cosmological nuisance parameters from the combinations of Planck 2018 mission of CMB observations, the Pantheon compilation of type Ia supernova observations, the cosmic chronometers observations for the Hubble parameter, the BAO observations, and the $f\sigma_8$ observations.

The mean value of $c_s^2$ is close to $1$, because the mean value of $\log _{\rm 10}(c_s^2)$ is close to zero which can be seen in Figures~\ref{fig:bounds_main} and~\ref{fig:prob_log10_cs2}. The higher values of $c_s^2$ (close to $1$) are loosely constrained i.e. it is allowed for large error bars. On the other hand, the lower values of $c_s^2$ are comparatively tightly constrained to lie far away from the mean value (in the aspect of the confidence interval). This can be seen in Figure~\ref{fig:prob_log10_cs2}. This means the homogeneous dark energy models are more favored than the clustering dark energy models with the recent cosmological observations.

The present work puts tighter constraints on parameters compared to similar earlier investigations \cite{Sergijenko:2014pwa,Kunz:2015oqa} by the use of more recent datasets.

Another interesting result, we find, is that the thawing behavior for the initial condition of the scalar field evolution is favorable at least for the polynomial and exponential potentials of the scalar field.

\section*{Data Availability Statement}
This manuscript has no associated data or the data will not be deposited. [Authors’ comment: This research uses publicly available cosmological data and wherever used cited properly.]

\begin{acknowledgements}
BRD would like to acknowledge IISER Kolkata for the financial support through the postdoctoral fellowship.
\end{acknowledgements}












\bibliographystyle{JHEP}
\bibliography{sample}

\end{document}